\documentclass[useAMS,usenatbib]{mn2e}
\usepackage{graphicx}
\usepackage{natbib}
\usepackage{amssymb}

\title[The emission in evolving FRI jets]{The evolution of the large-scale emission in FRI jets}
\author[P. Bordas, V. Bosch-Ramon and M. Perucho,]{P. Bordas$^{1,2}$\thanks{E-mail:
pol.bordas@uni-tuebingen.de}, V. Bosch-Ramon$^{3}$ and M. Perucho$^{4}$\\
$^{1}$Institut f\"ur Astronomie und Astrophysik, Universit\"at T\"ubingen, Sand 1, 72076 T\"ubingen, Germany\\
$^{2}${\textit{INTEGRAL}} Science Data Centre, Universit\'e de Gen\`eve, Chemin d'Ecogia 16, CH--1290 Versoix, Switzerland\\
$^{3}$Dublin Institute for Advanced Studies, 31 Fitzwilliam Place, Dublin 2, Ireland.\\
$^{4}$Departament d'Astronomia i Astrof\'{\i}sica. Universitat de Val\`encia. C/ Dr. Moliner 50, 46100 Burjassot (Val\`encia), Spain}
\begin{document}
\date{Released 2010 }

\pagerange{\pageref{firstpage}--\pageref{lastpage}} \pubyear{2010}

\def\LaTeX{L\kern-.36em\raise.3ex\hbox{a}\kern-.15em
    T\kern-.1667em\lower.7ex\hbox{E}\kern-.125emX}

\label{firstpage}

\maketitle

\begin{abstract} 

Recent observations in X-rays and gamma-rays of nearby FRI radio galaxies  have raised the question of the origin of the
emission detected in the termination structures of their jets. The study  of these structures can give information on the
conditions for particle acceleration and radiation at the front shocks. In addition, an evolutionary scenario can help to
disentangle the origin of the detected X-ray emission in young FRI sources,  like some Gigahertz Peaked Spectrum AGNs. This
work focuses  on the nature and detectability of the  radiation seen from the termination regions of evolving FRI jets. We
use the results of a relativistic, two-dimensional numerical  simulation of the propagation of an FRI jet, coupled with a
radiation model, to make predictions for the spectra and lightcurves of the thermal and non-thermal emission at different
stages of the FRI evolution. Our results show that under moderate magnetic fields, 
the synchrotron radiation would be the dominant
non-thermal channel, appearing extended in radio and more compact in X-rays, with relatively small flux variations with 
time.
The shocked jet synchrotron emission would dominate the X-ray band, although the shocked ISM/ICM thermal component alone may
be significant in old sources. 
Inverse Compton scattering of CMB photons could yield significant fluxes in the GeV and TeV
bands, with a non-negligible X-ray contribution. The IC radiation would present a bigger angular size in X-rays and GeV than
in TeV, with fluxes increasing with time.  We conclude that the thermal and non-thermal broadband emission from the
termination regions of FRI jets  could be detectable for sources located up to distances of a few 100~Mpc. 

\end{abstract}


\section{Introduction} \label{intro}

Extragalactic jets from Active Galactic Nuclei (AGN) inject energy in the interstellar and intracluster media (ISM
and ICM, respectively) at a rate between $\sim 10^{42}$ to  $\sim 10^{46}\,\rm{erg~s}^{-1}$, depending on  the source.
Fanaroff-Riley sources of type I \citep[FRI,][]{fr74} fall typically on the lower edge of this power spectrum. They show 
relativistic velocities at parsec-scales \citep{cg08} and disrupted structure at kiloparsec scales, whereas the more 
powerful FRII jets
keep collimated up to the medium interaction point, in which hot-spots can be observed at different frequencies. 
The
interaction of the jet with the ambient in FRI and FRII galaxies could be important to the extent that AGN feedback has
been claimed to be  a possible solution for the cooling flow problem via shock-heating or
mixing \citep[e.g.][]{qb01,zn05,mn07}. Also, this interaction can give rise to heating and particle acceleration via shocks, in which thermal and
non-thermal radiation is produced and can be used to study the properties of the flow and the medium. Following this idea,
\cite{hrb98} used a simple evolutionary model based on the work by \cite{rb97} to obtain the X-ray brightness of the thermal emission for
different initial jet properties. They claimed that only for dense enough cluster media, the count rates obtained would be
enough to detect this emission even for powerful FRII jets. \cite{kki07} have also derived estimates for the thermal MeV
emission from cocoons of radio galaxies depending on their age, with the result that only young cocoons, with ages $\ll
10^7\,\rm{yrs}$ could be detected at this energy band by present space-observatories. \cite{zn03} performed a series of
simulations of supersonic and underdense jets in a decreasing pressure atmosphere and showed that jets evolve in two
different phases regarding their high-energy thermal emission: a phase in which the shell formed by shocked material is
highly overpressured and radiative, and a later phase in which the shock is weaker and a deficit of X-ray emission is
expected from the lobes.

Regarding observations,  \cite{ka03} reported on the \textit{Chandra} detection of faint, extended X-ray emission from the 
jets and lobes of the radio galaxy 3C~15 (see also \citealt{harris06}). This emission is spatially correlated with that 
observed at 8.3 GHz radio frequencies \citep{leahy97}. The authors suggested that the same electron population responsible
for the  radio synchrotron emission upscatters the CMB photons to produce the diffuse X-ray radiation. \cite{sie08}  reported
the detection of X-ray emission from  Gigahertz Peaked Spectrum (GPS) and Compact Steep Spectrum (CSS) sources (13 quasars
and 3 galaxies, all of them powerful sources), and claimed that this radiation is most likely related to the accretion power
in all but one of the studied sources, in which the emission could be generated in the jet. They also discussed the
possibility that  the X-rays were produced in the bow shock formed by the expanding jet but found no evidence for this. 
\cite{kr03} and \cite{ckh07} reported detection of X-ray emission in Cen~A and NGC3801 using \emph{Chandra},  which was 
interpreted in terms of the bow shock driven by the injection of a jet. Modeling the emission as thermal, they obtained
bow-shock Mach numbers between 4 and 8. However, deeper observations of Cen~A \citep{kr07,cr09} showed that the emission from
a bow-shock region around the south-west lobe is better interpreted as of synchrotron origin, implying that the shock is
strong enough to accelerate particles up to Lorentz factors of $\sim 10^8$. In other sources, like the radio galaxy Fornax~A,
the lobes seem to emit non-thermal X-rays through inverse Compton (IC) of cosmic microwave background (CMB) photons
\citep[e.g.][]{fe95}, whereas the large-scale jet of M87 would be also a synchrotron emitter (e.g. \citealt{wi03}; see also
\citealt{ka05} for a discussion on extended jet emission and possible origins). Recently, the detection by {\it Fermi} of
extended GeV emission in the radio lobes of Cen~A \citep{ab10a}, likely via IC scattering of CMB photons, shows that
acceleration up to VHE is taking place in the disrupted jet region.

\cite{pm07} (PM07 hereafter) performed a simulation aimed to test the FRI jet evolution paradigm \citep{b84} and the model by
\cite{lb02} for the FRI jet of the radiogalaxy 3C~31. The simulation was done using a numerical code for relativistic
hydrodynamics, based on High-Resolution-Shock-Capturing schemes, to which it was added an equation of state that allows for a
specific treatment of two families of particles, leptons and baryons, and computes the adiabatic index in terms of the
composition of each cell.  The jet was injected in the numerical grid at $500\,\rm{pc}$ from the active nucleus, with
a radius of $60\,\rm{pc}$. The ambient  medium, composed by neutral hydrogen, has a profile in
pressure, density and temperature. 
Such a profile is required in most jet evolutionary models to account for the jet collimation at large distances. Furthermore, all models in which the jet is decelerated shortly after being ejected, within distances 1--10 kpc from the nucleus, require a gradient in the ambient pressure that permits the jet to prevent disruption due to external mass loading. This profile includes the contribution from a core region, dominant for distances up to $\sim 1.5$~kpc, and a more extended, hotter and less dense contribution from the galaxy group, which dominates at large distances. For a detailed discussion on the X-ray properties used to characterize the external medium in FR-I sources see, e.g., \cite{hr02} and \cite{lb02}.

The jet, leptonic in composition, was
injected with a velocity $v_{\rm j0}=0.87\,c$, density ratio with the ambient $\rho_{\rm j0}/\rho_{\rm a0}=10^{-5}$,
pressure ratio with the ambient $P_{\rm j0}/P_{\rm a0}\simeq 8$,  and temperature $4\times10^9\,\rm{K}$, resulting in a
kinetic luminosity $L_{\rm j}=10^{44}$~erg~s$^{-1}$. The simulated jet  evolved during $\approx 7\times 10^6\,\rm{yrs}$ up to
a distance of $15\,\rm{kpc}$. For further details on this simulation,  we refer the reader to PM07. 

In the present work, we use the results from this simulation of an FRI jet interacting with the ISM/ICM to compute the
produced thermal and non-thermal emission for different source ages. We have coupled a simplified radiation model for the cocoon
and the shell applied already to the context of microquasars \citep{br09} to the results of the simulations of PM07. In
this work, we make a specific use of the terms cocoon and shell. Namely, we refer to the cocoon as the region of the jet
shocked material, starting already at the recollimation (see below), and the shell as the region of shocked external
medium. We have also covered source ages older than $7\times 10^6\,\rm{yrs}$ using extrapolations of the main
hydrodynamical paratemers derived from the simulation results in PM07. In this way, we can make predictions for the
flux and the spectral evolution of the thermal (X-rays) and non-thermal (radio to gamma-rays) emission of an FRI jet for a
broad age range: $10^5-10^8$~yr. We discuss the relevance of the thermal and the non-thermal radiation, and the
possibility to produce high- (HE) and very high-energy (VHE) from the termination regions of FRI jets.

The paper has been organized as follows: in section~\ref{mod} we present the emission model and its results for different
stages of the cocoon and shell evolution, characterized using the simulations of PM07. The discussion of the results
and the conclusions are presented in Sect.~\ref{disc} and Sect.~\ref{concl}, respectively.

\section{Thermal/non-thermal emission from FRI jets}\label{mod}

\subsection{The model}

\begin{figure*}
\includegraphics[width=1\textwidth]{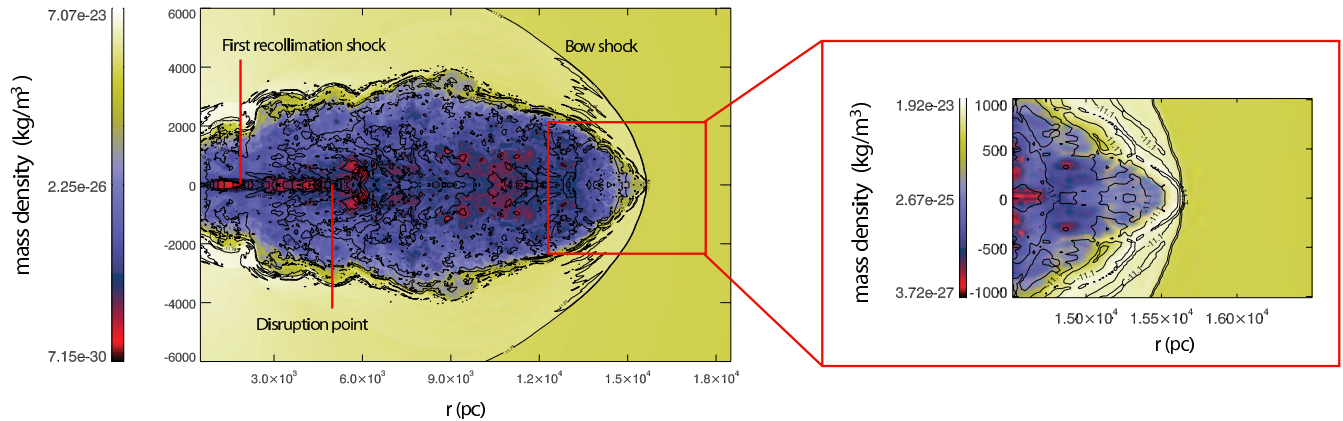}
\caption{Snapshot of the density in color scale for the simulation of an FRI jet after $7\times 10^6$~yr (for details, 
see PM07). The shell is clearly seen, as well as the recollimation shock, the disruption point 
and the turbulent cocoon region. The inset shows a zoom around the head of the bow shock. Pressure contours at the level of 2 and 8 $\times 10^{-12}$~erg~cm$^{-3}$ are labeled. The highest pressure zones correspond to the regions close to the  Rankine-Hugoniot conditions and are used to compute the $RH$ thermal emission, whereas the $Av$ contribution makes use of the values averaged over the whole shell (see text for details)}
\label{fig:sim1}
\end{figure*}

The model adopted here to study the non-thermal emission of the termination site of an FRI jet has been adapted from
\cite{br09}, in which the non-thermal radiation of a microquasar jet termination region was studied. In that paper, the
dynamics was based on the works by \cite{kaiser97} and \cite{falle91}, whereas here the dynamics has been extracted from the
simulations by PM07. The thermal emission has been computed using the information on the density and temperature obtained
from these simulations. For details on the properties of the jet at injection and the external medium, we address to Tables~1
and 2 in PM07 (see also Table 3 of the same work for a comparison with Cen~A and the radiogalaxy NGC~3801). Since in the
present case the jet is disrupted and a strong reverse shock is not produced (unlike in \citealt{br09}), the shock in the jet
(reverse shock) has not been considered. Instead, we have accounted for the strong recollimation shock as the particle
accelerator in the cocoon. For illustrative purposes, we present in Fig.~\ref{fig:sim1} a density map, with isobars, of the simulated jet
after $7\times 10^6$~yr of evolution  (see PM07). The inset in Fig.~\ref{fig:sim1} shows a zoomed view of the head of the
jet.

\subsubsection{Non-thermal particles and their emission}

Most of the accelerated electrons do not cool down significantly before overcoming the jet disruption point (DP) and reach
later on the turbulent cocoon. This is due to the fast motion downstream the recollimation shock, which efficiently carries
particles away along the jet axis down to the DP. Also, the compression of the shocked jet walls by the cocoon material
prevents expansion and therefore adiabatic cooling until electrons reach the DP. Beyond that point, relativistic electrons
spread in the cocoon via turbulent advection and diffusion. 

Since the material becomes trans- or subsonic after the recollimation shock, the pressure in the shocked jet, the cocoon and the shell 
should be relatively similar, the shell being denser but colder than the cocoon. In addition, the dominant photon field is the (homogeneous) CMB one, with radiation energy density $u_{\rm CMB}\approx 4.2 \times 10^{-13}\,(1 + z)^{4}$~erg~cm$^{-3}$ (we use $z=0$ in our calculations), provided that the emitter is located at a distance $\ga 2.6\,(L_{\rm nuc}/10^{43}\,{\rm erg~s}^{-1})$~kpc from the galaxy nucleus ($L_{\rm nuc}$ is the nucleus luminosity). This
allows us, at this stage, to simplify the cocoon region as an emitter with homogeneous properties (one zone) with the
recollimation shock as the injector of accelerated particles. The external medium shocked by the bow shock, i.e. the shell,
should be also mostly subsonic, and the same considerations regarding the photon field apply there. Therefore, we have also
adopted an homogeneous emitter approximation for the shell. 

In both the shell and the cocoon, the luminosity injected in the form of non-thermal particles has been taken as a 10\% of
the total jet kinetic luminosity, i.e. $L_{\rm nt}=0.1\,L_{\rm j}=10^{43}$~erg~s$^{-1}$. The magnetic field $B$ has been
fixed taking the magnetic energy density, $u_{\rm B}=B^2/8\pi$, to be 10\% of the ram/thermal pressure. Concerning particle
acceleration, the recollimation shock has been assumed to be relativistic, with an acceleration rate  $\dot{E}=\eta\,qBc$ with
$\eta=0.1$. For the bow shock, we have adopted the expression for a non-relativistic shock, in which
$\eta=\frac{1}{2\pi}(v_{\rm s}/c)^2$ (e.g. Drury 1983), where $v_{\rm bs}$ is the bow-shock velocity (typically here of $\sim
(1-2)\times 10^8$~cm~s$^{-1}$). These acceleration rates are to be compared to the synchrotron and IC loss rates (e.g. \citealt{blu70}) 
to derive
the maximum energy of electrons. The cooling timescales ($t_{\rm cool}=-E/\dot{E}_{\rm cool}$) of
synchrotron and (Thomson) IC processes are:  \begin{equation} t_{\rm syn}\approx 4\times 10^{12}\,(B/{\rm 10 \mu
G})^{-2}\,(E/{\rm 1 TeV})^{-1}\,{\rm s}\,,  \end{equation} and  \begin{equation} t_{\rm IC}\approx 1.6\times
10^{13}\,(u/10^{-12}{\rm erg~cm}^{-3})^{-1}\,(E/{\rm 1 TeV})^{-1}\,{\rm s}\,,  \end{equation} respectively, being $u$ the
total radiation energy density. An escape time has also to been considered since particles with enough energy would escape
the accelerator. This is derived by taking the gyroradius of the most energetic particles equal to the size of the
accelerator, i.e. the recollimation and bow-shock widths (Hillas 1984).  We do not consider the possible role of Fermi~II
stochastic or shear acceleration in the disrupted jet and cocoon regions (see, e.g., \citealt{ri07}; see also
\citealt{taylor09} for a deeper analysis of Fermi~II particle acceleration in the context of the Lobes of Cen~A), 
although these processes may be {\it absorbed} by our phenomenological treatment of the particle acceleration in the cocoon.

% ----------------------------------------------------
\begin{table*}
\begin{center}
\caption{Model parameters for the shell and cocoon for three different source ages used to compute the thermal and non-thermal emission.}
\label{tab:model}
\begin{tabular}{@{}lccc}
\hline
Parameter   & t$_{\rm src}=10^{5}$~yr & t$_{\rm src}=3 \times 10^{6}$~yr & t$_{\rm src}=10^{8}$~yr\\
\hline
%Jet power $L_{\rm jet}$ (erg~s$?{-1}$)  &  & $10^{44}$  & \\
%Non-thermal fraction $\xi$  &  & 0.1 & \\
%Equipartition fraction $\chi$ &  & 0.1 & \\
Bow shock velocity $v_{b}$ (c) & $8.7 \times 10^{-3}$ & $6.2 \times 10^{-3}$ & $4.4 \times 10^{-3}$ \\
Shell density $\rho_{\rm sh}$ (g~cm$^{-3}$) & $1.0 \times 10^{-24}$ & $4.0 \times 10^{-26}$ & $3.4 \times 10^{-27}$ \\
Shell temperature $T_{\rm sh}$ (K) & $8.3 \times 10^{7}$ & $2.7 \times 10^{7}$ & $3.5 \times 10^{6}$ \\
Shell radius $r_{\rm sh}$ (cm) & $1.6 \times 10^{20}$ & $6.5 \times 10^{21}$ & $1.2 \times 10^{23}$ \\
Shell and cocoon magnetic field $B$ (G) & $2.5 \times 10^{-4}$ & $3.1 \times 10^{-5}$ & $6.4 \times 10^{-6}$ \\

Cocoon radius $r_{\rm coc}$ (cm) & $5.5 \times 10^{20}$ & $2.2 \times 10^{20}$ & $4.0 \times 10^{20}$ \\
Shell maximum energy $E_{\rm max}^{\rm sh}$ (TeV) & 17.0 & 33.5 & 51.0 \\
Cocoon maximum energy $E_{\rm max}^{\rm coc}$ (TeV) & $1.5 \times 10^{3}$ & $4.4 \times 10^{3}$ & $9.1 \times 10^{3}$ \\
\hline
\end{tabular}

\medskip

\end{center}

\end{table*}
% ----------------------------------------------------

The properties of the non-thermal emitters in the cocoon and the shell are characterized by the ram/thermal pressure and the
bow-shock velocity (and the shock sizes when cooling is unefficient), which determine the magnetic field, the synchrotron
emission, indirectly the IC emission, and the acceleration efficiency. These conditions have been parameterized making use of
the results of the simulations of PM07 and their extrapolation to earlier and later times, covering an age range $t_{\rm
src}=10^5-10^8$~yr. We do not expect significant uncertainties from the extrapolations as long as the medium properties
present the same properties at larger distances than those covered by the simulated jet. Some of the model parameters
are listed in Table~\ref{tab:model} for both the shell and the cocoon regions.

The spectral aging of the non-thermal particle populations has been modeled considering the evolution of the physical
conditions in each interaction region. The particle energy distribution at a given time, $N(E,t_{\rm src})$, is calculated by
adding the different evolved injected populations, $Q(E,t)$ ($\propto E^{-p}$), from $t=0$ up to $t_{\rm src}$. The time
resolution of particle injection is $\Delta t\la t_{\rm cool} (t)$. Maximum particle energies, $E_{\rm max}(t)$, are also
computed for each time step due to the time dependence of the magnetic field, the accelerator size and the shock velocity. 
For simplicity, a spectral index $p=2.1$ has been used in our calculations for both the recollimation and the bow shock. We
note that, together with synchrotron and IC cooling, the expansion of the jet termination structure introduces an adiabatic
loss timescale \citep[see][]{br09} $\sim l/v_{\rm b}\approx (5/3)\,t_{\rm src}$, where $l$ is the size of the whole structure. In
addition to synchrotron and IC processes, relativistic Bremsstrahlung \citep{blu70} could also take place in the shell, and
protons may be accelerated and eventually could interact with the shocked jet medium through proton-proton ($pp$) collisions
(see \citealt{kelner06}). However, the densities $n$ of targets for relativistic Bremsstrahlung and $pp$ emission in the
shell, the largest in the jet termination region, are low, and the cooling timescales:  

\begin{equation} t_{\rm rel.br/pp}\sim 10^{18}\,(n/10^{-3}\,{\rm cm}^{-3})\,{\rm s}\gg t_{\rm src}\,. \end{equation} 

This implies radiation efficiencies much smaller than those of synchrotron and IC. In addition, it is worth noting that synchrotron proton emission \citep[e.g.][]{aha00}, under equipartition magnetic fields, could overcome IC radiation around 100~MeV, although only for very young sources ($\ll 10^{5}$~yr) this component may be significant. We do not further consider either relativistic Bremsstrahlung or proton radiation processes in this work.

\subsubsection{Thermal emission}\label{themmod}

The thermal emission has been computed making use of the simulation results and their extrapolation to the  $t_{\rm
src}$-range considered here. Given the strong density dependence of thermal Bremsstrahlung, we have only accounted for the
contribution from the shell, much denser than the cocoon. Furthermore, we have simplified the calculations of the thermal
radiation as it would be coming from two regions (see Fig.~\ref{fig:sim1}). One, cooler (ultraviolet -UV-/soft X-rays) but
brighter, corresponds to the averaged shell conditions ($Av$), and another one, fainter but hotter (hard X-rays), corresponds
to a region close to the apex of the bow shock ($RH$), in which the shell has properties close to those given by the jump
conditions of Rankine-Hugoniot. The volume of the latter region is about $3-4$\% of that of the whole shell (see inset in
Fig.~\ref{fig:sim1}), which corresponds to the volume limited by the isobars satisfying $P \gtrsim \frac{1}{2}\, P_{\rm R-H}$, where $P_{R-H}$ is the shell pressure right behind the bow shock and corresponds to the Rankine-Hugoniot jump conditions. Given the high temperatures in the shell, we have calculated  the thermal Bremsstrahlung assuming that the plasma is fully ionized, with electrons and protons in equipartition. At this stage, we have not considered line emission.

\subsection{Results}

We have studied the thermal and non-thermal emission produced in the shell and the cocoon separately. We consider here the
contribution from a single FRI jet, so the predicted luminosities should be scaled by a factor of two to obtain the whole
source emission under similar ambient conditions for jet and counter-jet.

\subsubsection{Non-thermal emission}

\begin{figure*}
\includegraphics[width=1\textwidth]{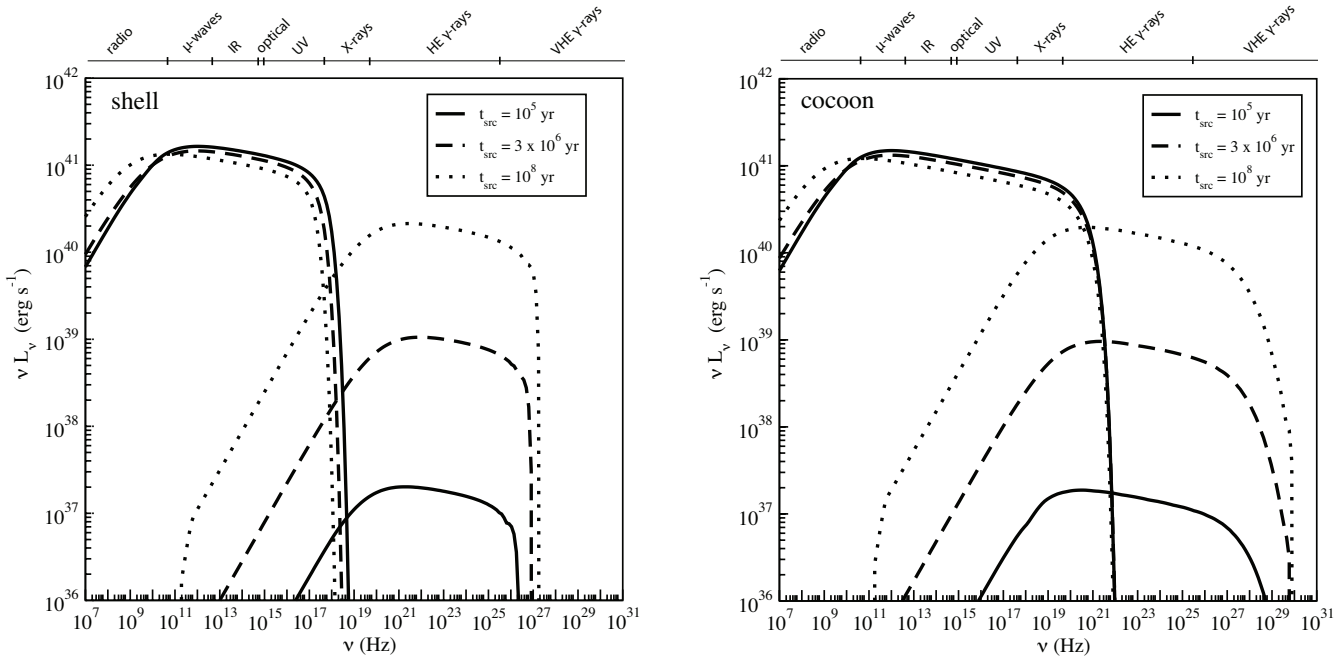}
\caption{Computed SEDs of the non-thermal synchrotron and IC emission from the shell and the cocoon for three different ages: $10^5$ (solid line), $3\times 10^6$ (long-dashed line) and $10^8$~yr (dotted line). We have accounted for the contribution of one jet/medium interaction region only, so the values here should be scaled by a factor of two to get the whole source emission.}
\label{nonthermal_SED}
\end{figure*}

The bow shock can accelerate electrons up to energies $E_{\rm max}\approx 17~$~TeV ($10^{5}$~yr) to $51$~TeV at
($10^{8}$~yr), and is limited by synchrotron losses at any time. This trend of higher $E_{\rm max}$ for older sources comes
from the energy gain to loss ratio $\propto B^{-1}$ under synchrotron dominance. Since $v_{\rm b}$ decreases moderately, from
$\approx 2.8 \times 10^{8}$ to $1.3 \times 10^{8}$~cm~s$^{-1}$, the strongest variation in the shell acceleration rate comes
from the $B$-evolution, which goes from $\approx 2.5\times 10^{-4}$ ($10^{5}$~yr) to $6.4\times 10^{-6}$~G ($10^{8}$~yr). The
shell IC emission is dominated by scatterings with CMB photons. In the cocoon region, we have assumed the recollimation shock
to be the accelerator site. The maximum energy also grows here, going from $\approx 1.5 \times 10^3$ to $9\times 10^{3}$~TeV.
Since $u_{\rm B}$ is proportional to the pressure and the latter is similar in the cocoon and the shell (see Figs.~5 and 6 in
PM07), $B$ is also similar in both regions. As in the shell, synchrotron losses dominate for the magnetic field strengths and
ages considered here. The high values of $E_{\rm max}$ in the recollimation shock are expected since the acceleration rate is
assumed to be $\sim (c/v_{\rm s})^{2}$ times more efficient here than in the non-relativistic bow shock. The large distance
of the recollimation shock to the galaxy nucleus makes the CMB IC to dominate over other IC components, although for very
young sources the galaxy nucleus could be relevant.

The non-thermal spectral energy distributions (SED) for the cocoon and the shell, at $t_{\rm src}=10^{5}$, $3 \times 10^{6}$
and $10^{8}$~yr, are shown in Fig.~\ref{nonthermal_SED}. The obtained radio and X-ray synchrotron luminosities in
both regions are at the level of $2\times 10^{41}$~erg~s$^{-1}$. The approximate constancy of the luminosities with time is
due to the fact that particles have reached the steady state at $t_{\rm src}$ through synchrotron cooling\footnote{Actually,
the adiabatic cooling, as approximated here, takes $\sim 1/2$ of the particle energy after a time $\sim t_{\rm src}$, the
rest of the energy going to radiation.} and the assumed constancy of $L_{\rm nt}$. The decrease of $B$
with time, and therefore the growth of $t_{\rm syn}$, is compensated by the increase of time available for cooling. The
synchrotron break frequency, corresponding to the electron energy at which $t_{\rm syn}(E)\approx t_{\rm src}$, and the
highest synchrotron frequency, $\nu_{\rm syn~max}\propto B\,E_{\rm max}^2$, are shifted down for older sources. The former 
effect makes the radio luminosity to increase at the late stages of the evolution of both the cocoon and the shell, whereas
the latter decreases the X-ray luminosity in the shell due to the decrease of $\nu_{\rm syn~max}$ with time. The slightly
different conditions in the shell yield a higher break frequency, which implies a factor  $\sim 2$ lower radio luminosity in
this region compared to that of the cocoon. 

The IC luminosity grows as long as this process becomes more efficient compared to synchrotron and adiabatic cooling,
which is shown by the decrease of $u_{B}/u_{\rm rad}$ from $\approx 5 \times 10^{3}$ ($10^5$~yr) to 4 ($10^{8}$~yr). As expected
from the $E_{\rm max}$-values given above and the similar energy budget, the cocoon and the shell have similar HE
luminosities, but the cocoon is few times brighter at VHE than the shell due to its much higher maximum frequency. In both
regions the bolometric IC luminosities grow similarly with time, reaching $\sim 10^{42}$ and $10^{41}$~erg~s$^{-1}$  at HE
and VHE, respectively. 

\begin{figure*}
\includegraphics[width=0.9\textwidth]{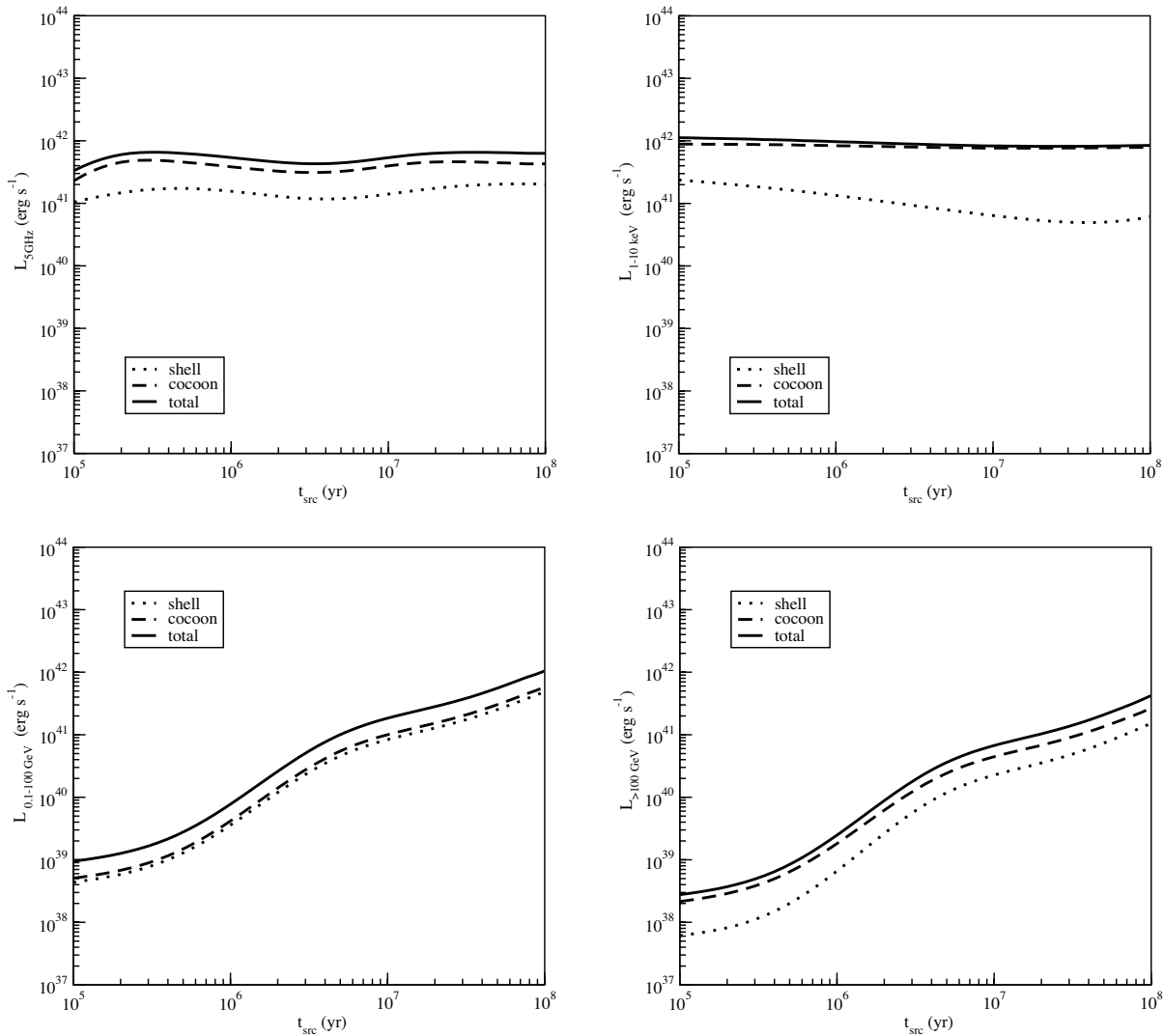}
\caption{Computed non-thermal lightcurves of the radio (${\rm 5~GHz}\times L_{\rm 5~GHz}$, top-left), 
X-ray (bolometric: 1--10~keV, top-right) and gamma-ray emission 
(bolometric: 0.1--100~GeV, bottom-left; bolometric: $>100$~GeV: bottom-right) in the age range $t_{\rm src}=10^5-10^8$~yr. 
The lightcurves for the cocoon and the bow-shock emission, and the summation of both, are shown.}
\label{nonthermal_lc}
\end{figure*}

The lightcurves for the luminosities in radio (${\rm 5~GHz}\times L_{\rm 5~GHz}$), X-rays (bolometric: 1--10~keV), HE
(bolometric: 0.1-100~GeV) and VHE (bolometric: 0.1--100~TeV), for both the cocoon and the shell, are presented in
Fig.~\ref{nonthermal_lc}. The lightcurves show in more detail the time behavior of the non-thermal radiation at different
wavelengths discussed above. The complex and smooth shape of the lightcurves, most clear for the HE and the VHE
emission, is a consequence of the complex hydrodynamical evolution of the whole interaction structure propagating in an
inhomogeneous external medium.

\subsubsection{Thermal emission}

\begin{figure} 
\includegraphics[width=0.45\textwidth]{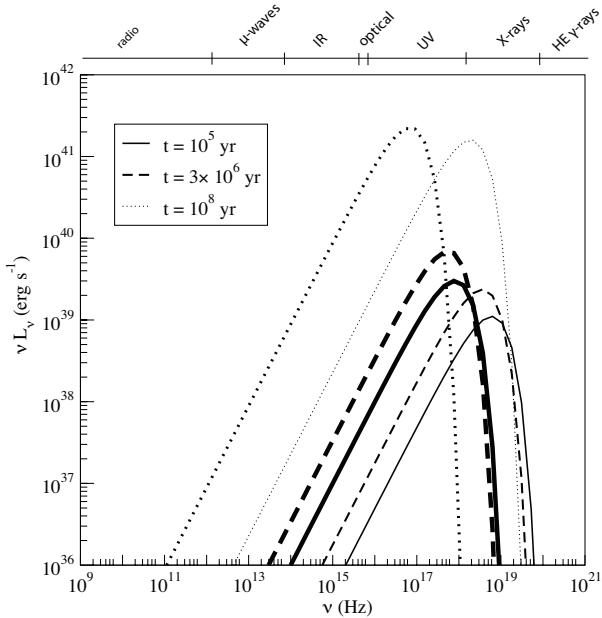} \caption{Computed SEDs of the shell
thermal emission for three different ages: $10^5$ (solid line), $3\times 10^6$ (long-dashed line) and $10^8$~yr 
(dotted line). 
The two components are shown, one
corresponding to the whole shell (thick lines), 
and another one related to a shell region with conditions similar to those of
Rankine-Hugoniot (thin lines).} 
\label{thermal_SED} 
\end{figure}

As mentioned in Sect.~\ref{themmod}, thermal Bremsstrahlung is also expected from the shell. Figure~\ref{thermal_SED} shows
the three SEDs ($t_{\rm src}=10^{5}$, $3 \times 10^{6}$ and $10^{8}$~yr) computed adopting a simplified model for the thermal
emitter of the shell considering the shell averaged values, $Av$, and the conditions right behind the bow shock apex, $RH$.
The slowdown of the bow shock and the velocity dependence of the postshock temperature, $\propto v_{\rm bs}^2$, leads to a
decrease in the peak of the thermal emission with time, whereas the increase of the shell mass yields higher thermal
bolometric luminosities as the source gets older. For the age range $t_{\rm src}=10^5-10^8$~yr, the thermal luminosities go
from $10^{39}$ to few times $10^{41}$~erg~s$^{-1}$, with the shell  and the hot postshock region components peaking from soft
X-rays to UV and from hard to soft X-rays, respectively. The hot postshock region dominates the SED in hard X-rays by a
factor of a few over the shell thermal and both shell and cocoon non-thermal components for $t_{\rm src}=10^8$~yr. 

\begin{figure}
\includegraphics[width=0.45\textwidth]{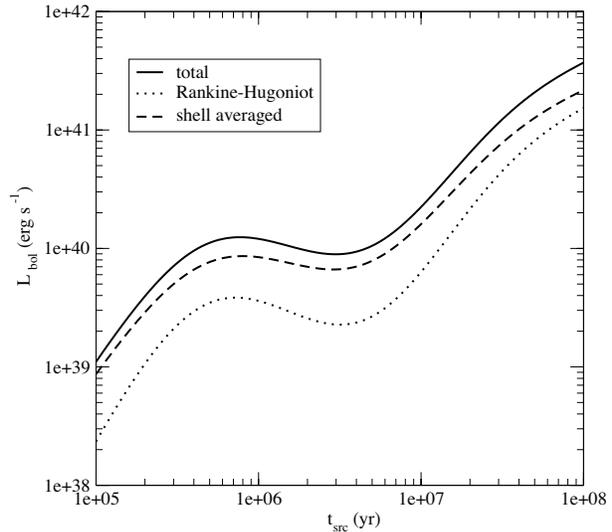}
\caption{Evolution of the computed thermal bolometric luminosity with time. Two components are shown, one computed with 
density and temperature averaged in the whole shell, $Av$ (dotted line), and the other one corresponding to a region with 
conditions similar to those of a strong shock, $RH$ (long-dashed line).
The summation of both components is also shown (thick solid line).}
\label{thermal_lc}
\end{figure}

Figure~\ref{thermal_lc} shows the thermal lightcurves (bolometric) for $t_{\rm src} = 10^{5}-10^{8}$~yr. Thermal
Bremsstrahlung increases from $t_{\rm src}=10^{5}$ to $\sim 10^{6}$~yr, when it shows a relative maximum. Then the luminosity
slightly decreases until $t_{\rm src}\sim 3\times 10^{6}$~yr, time in which there is a transition in the external medium,
from the denser galaxy core to the rarefied galaxy group medium (see PM07 for details). Later, the emission increases again.
The component $Av$ dominates the thermal bolometric luminosity in young sources, but the component $RH$ becomes similarly
bright at $t_{\rm src}\sim 10^{8}$~yr.

\section{Discussion}\label{disc}

\subsection{Radio}

The cocoon and the bow shock show both a similar pattern of their non-thermal radio emission, although a higher $E_{\rm max}$
makes the cocoon emission to extend to higher energies. The accumulation and aging of the injected particles lead to, for
$t_{\rm src}=10^8$~yr, a break in the synchrotron spectrum around the radio frequencies. The cocoon would be the dominant
radio emitter, with fluxes as high as $\sim 10^{-12}\,(d/100~$Mpc$)^{-2}$~erg~cm$^{-2}$~s$^{-1}$ or $\sim 10$~Jy at 5~GHz
from a region of few times $10'\,(d/100~$Mpc$)^{-1}$ angular size. The spectral index would appear inverted due to particle
aging, with $\alpha\sim 1$ ($F_\nu\propto \nu^{-\alpha}$). The properties of the radio emission from the interaction
jet-medium structure are comparable with those observed for instance in 3C~31, with radio luminosities at 4.75~GHz of about
$3\times 10^{40}$~erg/s  \citep{a92}, or with the ones of 3c~15, in which fluxes of a few $\times
10^{-14}$~erg~cm$^{-2}$~s$^{-1}$ ($d\sim$~300~Mpc) are found \citep{ka03}. The predicted shell radio flux is sligthly below
the cocoon one, although effects of limb brightening may enhance the detectability of the former. The radio lightcurve is
quite steady, with some small variations. Despite the fact that the magnetic field gets weaker with time, the accumulation of
radio emitting electrons compensates it, and the final emission keeps roughly constant.

\subsection{X-rays}

Thermal X-rays are produced in the shell, with a temporal evolution smoother and more complex than in the case of a
homogeneous medium. The existence of different emitting regions in the shell would lead to a relatively flat thermal spectrum
in X-rays, with a bolometric flux from $\sim 10^{-15}\,(d/100~$Mpc$)^{-2}$ ($10^5$~yr) to few times
$10^{-13}\,(d/100~$Mpc$)^{-2}$~erg~cm$^{-2}$~s$^{-1}$ ($10^8$~yr). We note that the thermal emission would be restricted to
different angular size regions depending on the photon energy. The hard X-ray photons would come from the apex of the bow
shock, with typical angular size of a few $1'\,(d/100~$Mpc$)^{-1}$ (for $t_{\rm src}\sim 10^8$~yr), and the lower energies
would be dominated by the whole cooler shell emission, with an angular size of a few $10'\,(d/100~$Mpc$)^{-1}$. Limb
brightening effects could play a role, showing a thin structure along the limb of the shell with the hottest region at the
apex. It is worth noting that under the adopted $L_{\rm nt}$-value and $B$-equipartition fraction, the shell thermal emission
dominates the emission except in hard X-rays. We also note that given the moderate velocities of the bow shock, thermal
photons cannot reach energies as high as those discussed in \cite{kki07}. Nevertheless, for the shell properties considered in this work, the thermal cooling time-scale $t_{\rm th} \sim 2.5 \times 10^{9} (\frac{T}{10^{7} {\rm K}})^{0.5} (\frac{n_{e}}{10^{-2} {\rm cm}^{-3}})^{-1}$~yr is greater than $t_{\rm src}$. Assuming that the bow shock keeps being adiabatic and strong all along the source age, and hence not displaying a transition to a much weaker shock regime (see e.g. Zanni et al. 2003), the thermal bolometric luminosity increases with time.

Regarding non-thermal X-rays, the dominant emission comes also from the cocoon, with fluxes $\sim
10^{-13}\,(d/100~$Mpc$)^{-2}$~erg~cm$^{-2}$~s$^{-1}$, although again limb brightening effects may increase the shell
detectability. In fact, in the case of Cen~A, the shell seems to be the dominant source of non-thermal X-rays (Croston et al.
2009). This difference could be explained by a higher $E_{\rm max}$ in the shell of that source. In addition, a relatively
recent decrease in jet power would have affected first the cocoon synchrotron emission, making this radiation fainter while
the shell emission would remain at similar levels for a time $\ga 10\,{\rm kpc}/c\sim 3\times 10^4$~yr. The lifetime of X-ray
synchrotron electrons, $\sim 10^{11}\,{\rm s}$, is much shorter than in radio, and $\ll t_{\rm src}$ as well. This implies
that these particles may not have time to reach the whole emitting structures, and their radiation may come mostly from the
inner regions of the cocoon or the bow-shock apex. Note that this may lead to a violation of the assumption of an homogeneous
emitter. This X-ray synchrotron emission concentrated around the recollimation shock is compatible with the large-scale
jet X-ray emission found in 3C~31 by \cite{hr02}. If a strong recollimation shock is indeed the origin of these large-scale
jet X-rays, then the hypothesis that jet disruption in 3C~31 is caused by shock triggered instabilities is favored against
stellar wind mass-load (as proposed by \citealt{lb02}; see also PM07).

Like the radio emission from the cocoon, particle aging makes the X-ray synchrotron spectrum in the cocoon and the shell to
be relatively soft, with photon index $\Gamma\ga 2$ ($n_\nu\propto \nu^{-\Gamma}$). The total non-thermal X-ray flux is
roughly constant for the explored range of $t_{\rm src}$, although the shell contribution decreases significantly with time.
Close to the recollimation shock, before reaching the DP, the spectrum in soft X-rays would appear harder, since the
corresponding emitting electrons could not have time to cool. We note that X-ray fluxes at ages $\sim 10^6-10^7$~yr would not
be very different from those found in young sources by \cite{sie08}. Also, the non-thermal X-ray fluxes $\sim
10^{-14}$~erg~cm$^{-2}$~s$^{-1}$ of 3C 15 reported by \cite{ka03} (see Fig.~8 in that work) imply a non-thermal luminosity of
$\approx 1.3 \times 10^{41}$~erg~s$^{-1}$ at 300~Mpc, in good agreement with the values predicted here.

As noted in Sect.~\ref{intro}, the extended emission in some FRI jets has been interpreted as IC instead of synchrotron
emission. As shown for instance in Fig.~5 of \cite{is06} for Fornax~A, in some cases the X-ray emission is difficult to
reconcile with a synchrotron origin. We note however that the
predicted IC X-ray spectra (Fig.~\ref{nonthermal_SED}) would appear similar to the one shown in \cite{is06}. Therefore, for
sources older than those considered here and/or lower $B$-values, in which synchrotron emission would be less relevant, cocoon and
shell IC would dominate the X-ray output (covering roughly the whole cocoon/shell structure given the long cooling timescales
of IC X-ray electrons). We remind that the complex medium of the turbulent cocoon region, not treated here, may enhance the
X-rays in certain compact regions. Much lower surface brightness could prevent the detection of the more diffuse X-rays
coming from larger regions of the cocoon. X-ray emitting electrons may also lose a significant fraction of their energy right
after the recollimation shock. Given the transrelativistic velocities in that region, Doppler boosting would beam the
emission favoring certain viewing angles. This effect has not been accounted for in the present study. 

Finally, we remark that thermal emission seems unavoidable given the medium densities and bow-shock temperatures, whereas IC
depends on $L_{\rm nt}$, as well as synchrotron, which also depends on $B$, none of these two  parameters being well
determined.

\subsection{Gamma-rays}

The predicted SEDs in the HE-VHE range are similar for both the cocoon and the shell, although the latter shows a lower
maximum photon energy. The HE SED is close to flat, and becomes steeper at VHE. We have not accounted for EBL gamma-ray
absorption, which would become significant at distances larger than 100~Mpc. Regarding the lightcurve, the emission increases
with time mainly due to the increasing efficiency of the CMB IC channel as  compared with synchrotron and adiabatic losses.
We note that the gamma-ray fluxes for a source with $t_{\rm src}\sim 10^8$~yr are around $\sim 10^{-12}\,(d/100\,{\rm
Mpc})$~erg~cm$^{-2}$~s$^{-1}$. At HE, such a source may require very long exposures to be detected by, e.g., {\it Fermi},
although it cannot be discarded that very nearby sources, or sources with bigger non-thermal efficiencies or jet powers,
could be detected after few years of observations. Actually, {\it Fermi} has already detected several FRI galaxies up to few
hundred Mpc distances \citep{ab10b}, including the extended radio lobes of Cen~A (at a distance $\sim 4$~Mpc;
\citealt{is98}), presenting fluxes similar to those predicted here. At VHE, the fluxes would be detectable by the current
instruments, although the extension of the source, of tens of arcminute at 100~Mpc, and the steepness of the spectrum above
$\sim 100$~GeV, may make a detection difficult. In the case of Cen~A, detected by HESS \citep{ah09}, the emission seems to
come only from the core, but this is expected given the large angular size of the lobes of this source, which would dilute
its surface brightness too much. In general, long exposures with present Cherenkov instruments, like HESS, MAGIC II and
VERITAS, and the forthcoming CTA, may allow the detection of VHE emission from FRI jet lobes, and possibly carry out
energy-dependent morphological studies.

\section{Conclusions}\label{concl}

We have applied a radiative model to a prototypical FRI jet characterizing the flow with the results of hydrodynamical
simulations. Thermal Bremsstrahlung (X-rays), and non-thermal synchrotron (radio-X-rays) and CMB IC (X-rays-gamma-rays) have
been considered as the emission mechanisms. 

From our study we conclude that, for moderate non-thermal luminosities, radio
lobes of FRI radio galaxies are good candidates to be detected in the whole spectral range, with the radiation appearing
extended in most of the energy bands. The precise extension of the emitting regions is hard to calculate, and may depend e.g. on the source distance, the instrument resolution and the capability to disentangle the non-thermal emission from the background contribution. Our study does not aim to
provide specific values of the source extension at different energies, and offers only rough estimates of the overall emission morphology.

Our results show that soft X-rays may be likely dominated by synchrotron emission up to ages $\sim 10^8$~yr,
with IC tending to be dominant for older sources. Thermal X-rays seem unavoidable and may dominate in hard X-rays in old
sources even if a non-thermal component is present. The low surface brightness may require long observation times for the
detection in X- and gamma-rays, although the steady nature of these sources can help in this regard. Moderate resolution
radio and X-ray data, with long enough exposures, can allow the direct comparison between predictions of simulations and
observational data, thus giving clues on the hydrodynamics of the present flows and their surroundings. Also, any nearby
galaxy of this kind can be a suitable candidate for an eventual gamma-ray detection. Non-thermal synchrotron X-rays and HE
and VHE gamma-rays provide suitable information to study particle acceleration in the jet termination regions.

\section*{Acknowledgments}
PB acknowledges support from the German Federal Ministry of Economics and Technology through DLR grant 50 OG 1001. PB also 
acknowledges the excellent work conditions at the \textit{INTEGRAL} Science Data Center. MP acknowledges support from a ``Juan de la Cierva'' contract of the Spanish 
``Ministerio de Ciencia y Tecnolog\'{\i}a'' and by
the Spanish ``Ministerio de Educaci\'on y Ciencia'' and the European Fund for Regional Development 
through grants AYA2007-67627-C03-01 and AYA2007-67752-C03-02 and Consolider-Ingenio 2010, ref. 20811. 
V.B-R. acknowledges support of the Spanish MICINN under grant
AYA2007-68034-C03-1 and FE\-DER funds.
V.B-R. also acknowledges the support of the European Community under a Marie Curie Intra-European 
fellowship.

\label{lastpage}

\end{document}